\documentclass[longauth]{aa}
\usepackage{graphicx}
\usepackage{txfonts}
\usepackage{xcolor}
\usepackage{hyperref}
\hypersetup{colorlinks, linkcolor=blue, citecolor=blue, urlcolor=blue}
\newcommand{\pivec}{\mbox{\boldmath $\pi$}}
\newcommand{\muvec}{\mbox{\boldmath $\mu$}}

\newcommand{\te}{t_{\rm E}}
\newcommand{\thetae}{\theta_{\rm E}}

\newcommand{\pie}{\pi_{\rm E}}
\newcommand{\pien}{\pi_{{\rm E},N}}
\newcommand{\piee}{\pi_{{\rm E},E}}
\newcommand{\dl}{D_{\rm L}}
\newcommand{\ds}{D_{\rm S}}

\definecolor{brown}{rgb}{0.59, 0.29, 0.0}
\definecolor{darkgreen}{rgb}{0.0, 0.42, 0.24}
\definecolor{darkblue}{rgb}{0.01, 0.31, 0.59}
\definecolor{darkblue}{rgb}{0.0, 0.25, 0.42}
\definecolor{blue}{rgb}{0.0,0.0,1.0}
\definecolor{green}{rgb}{0.0,1.0,0.0}

\def\eqalign#1{\null\,\vcenter{\openup\jot
        \ialign{\strut\hfil$\displaystyle{##}$&$
        \displaystyle{{}##}$\hfil \crcr#1\crcr}}\,}
\begin{document}

\title{KMT-2021-BLG-1150Lb: Microlensing planet detected through a densely covered planetary-caustic signal}
\titlerunning{KMT-2021-BLG-1150Lb: Microlensing planet detected through planetary caustic}

\author{
% leading author -----------------------------
     Cheongho~Han\inst{01}
\and Youn~Kil~Jung\inst{02,03}
\and Ian~A.~Bond\inst{04}
\and Andrew~Gould\inst{05,06}
\\
(Leading authors)\\
% KMTNet ---------------------------
     Sun-Ju~Chung\inst{02,07}
\and Michael~D.~Albrow\inst{08}
\and Kyu-Ha~Hwang\inst{02}
\and Yoon-Hyun~Ryu\inst{02}
\and In-Gu~Shin\inst{07}
\and Yossi~Shvartzvald\inst{09}
\and Hongjing~Yang\inst{10}
\and Jennifer~C.~Yee\inst{07}
\and Weicheng~Zang\inst{07,10}
\and Sang-Mok~Cha\inst{02,11}
\and Doeon~Kim\inst{01}
\and Dong-Jin~Kim\inst{02}
\and Seung-Lee~Kim\inst{02}
\and Chung-Uk~Lee\inst{02}
\and Dong-Joo~Lee\inst{02}
\and Yongseok~Lee\inst{02,11}
\and Byeong-Gon~Park\inst{02}
\and Richard~W.~Pogge\inst{06}
\\
(The KMTNet collaboration)\\
%% MOA ---------------------------
     Fumio~Abe\inst{12}
\and Richard~Barry\inst{13}
\and David~P.~Bennett\inst{13,14}
\and Aparna~Bhattacharya\inst{13,14}
\and Hirosame~Fujii\inst{12}
\and Akihiko~Fukui\inst{15,16}
\and Ryusei~Hamada\inst{17}
\and Yuki~Hirao\inst{17}
\and Stela~Ishitani Silva\inst{13,18}
\and Yoshitaka~Itow\inst{12}
\and Rintaro~Kirikawa\inst{17}
\and Iona~Kondo\inst{17}
\and Naoki~Koshimoto\inst{19}
\and Yutaka~Matsubara\inst{12}
\and Sho~Matsumoto\inst{17}
\and Shota~Miyazaki\inst{17}
\and Yasushi~Muraki\inst{12}
\and Arisa~Okamura\inst{17}
\and Greg~Olmschenk\inst{13}
\and Cl{\'e}ment~Ranc\inst{20}
\and Nicholas~J.~Rattenbury\inst{21}
\and Yuki~Satoh\inst{17}
\and Takahiro~Sumi\inst{17}
\and Daisuke~Suzuki\inst{17}
\and Taiga~Toda\inst{17}
\and Mio~Tomoyoshi\inst{17}
\and Paul~J.~Tristram\inst{22}
\and Aikaterini~Vandorou\inst{13,14}
\and Hibiki~Yama\inst{17}
\and Kansuke~Yamashita\inst{17}
\\
(The MOA Collaboration)\\
}

\institute{
      Department of Physics, Chungbuk National University, Cheongju 28644, Republic of Korea,                                                            % (01)
\and  Korea Astronomy and Space Science Institute, Daejon 34055, Republic of Korea                                                                       % (02)
\and  Korea University of Science and Technology, Korea, (UST), 217 Gajeong-ro, Yuseong-gu, Daejeon, 34113, Republic of Korea                            % (03)
\and  Institute of Natural and Mathematical Science, Massey University, Auckland 0745, New Zealand                                                       % (04)
\and  Max-Planck-Institute for Astronomy, K\"{o}nigstuhl 17, 69117 Heidelberg, Germany                                                                   % (05)
\and  Department of Astronomy, Ohio State University, 140 W. 18th Ave., Columbus, OH 43210, USA                                                          % (06)
\and  Center for Astrophysics $|$ Harvard \& Smithsonian, 60 Garden St., Cambridge, MA 02138, USA                                                        % (07)
\and  University of Canterbury, Department of Physics and Astronomy, Private Bag 4800, Christchurch 8020, New Zealand                                    % (08)
\and  Department of Particle Physics and Astrophysics, Weizmann Institute of Science, Rehovot 76100, Israel                                              % (09)
\and  Department of Astronomy, Tsinghua University, Beijing 100084, China                                                                                % (10)
\and  School of Space Research, Kyung Hee University, Yongin, Kyeonggi 17104, Republic of Korea                                                          % (11)
\and  Institute for Space-Earth Environmental Research, Nagoya University, Nagoya 464-8601, Japan                                                        % (12)
\and  Code 667, NASA Goddard Space Flight Center, Greenbelt, MD 20771, USA                                                                               % (13)
\and  Department of Astronomy, University of Maryland, College Park, MD 20742, USA                                                                       % (14)
\and  Department of Earth and Planetary Science, Graduate School of Science, The University of Tokyo, 7-3-1 Hongo, Bunkyo-ku, Tokyo 113-0033, Japan      % (15)
\and  Instituto de Astrof{\'i}sica de Canarias, V{\'i}a L{\'a}ctea s/n, E-38205 La Laguna, Tenerife, Spain                                               % (16)
\and  Department of Earth and Space Science, Graduate School of Science, Osaka University, Toyonaka, Osaka 560-0043, Japan                               % (17)
\and  Department of Physics, The Catholic University of America, Washington, DC 20064, USA                                                               % (18)
\and  Department of Astronomy, Graduate School of Science, The University of Tokyo, 7-3-1 Hongo, Bunkyo-ku, Tokyo 113-0033, Japan                        % (19)
\and  Sorbonne Universit\'e, CNRS, UMR 7095, Institut d'Astrophysique de Paris, 98 bis bd Arago, 75014 Paris, France                                     % (20)
\and  Department of Physics, University of Auckland, Private Bag 92019, Auckland, New Zealand                                                            % (21)
\and  University of Canterbury Mt.~John Observatory, P.O. Box 56, Lake Tekapo 8770, New Zealand                                                          % (22)
}

%\date{Received ; accepted}

% \abstract{}{}{}{}{} 
% 5 {} token are mandatory
\abstract
% context heading (optional)
% {} leave it empty if necessary  
{}
% aims heading (mandatory)
{
Recently, there have been reports of various types of degeneracies in the interpretation 
of planetary signals induced by planetary caustics.  In this work, we check whether such 
degeneracies persist in the case of well-covered signals by analyzing the lensing event 
KMT-2021-BLG-1150, for which the light curve exhibits a densely and continuously covered 
short-term anomaly.
}
% methods heading (mandatory)
{
In order to identify degenerate solutions, we thoroughly investigate the parameter space 
by conducting dense grid searches for the lensing parameters. We then check the severity 
of the degeneracy among the identified solutions.
}
% results heading (mandatory)
{
We identify a pair of planetary solutions resulting from the well-known inner-outer
degeneracy, and find that interpreting the anomaly is not subject to any degeneracy 
other than the inner-outer degeneracy. The measured parameters of the planet separation
(normalized to the Einstein radius) and mass ratio between the lens components are 
$(s, q)_{\rm in}\sim (1.297, 1.10\times 10^{-3})$ for the inner solution and 
$(s, q)_{\rm out}\sim (1.242, 1.15\times 10^{-3})$ for the outer solution. According to 
a Bayesian estimation, the lens is a planetary system consisting of a planet with a mass 
$M_{\rm p}=0.88^{+0.38}_{-0.36}~M_{\rm J}$ and its host with a mass 
$M_{\rm h}=0.73^{+0.32}_{-0.30}~M_\odot$ lying toward the Galactic center at a distance 
$D_{\rm L} =3.8^{+1.3}_{-1.2}$~kpc.  By conducting analyses using mock data sets prepared 
to mimic those obtained with data gaps and under various observational cadences, it is 
found that gaps in data can result in various degenerate solutions, while the observational 
cadence does not pose a serious degeneracy problem as long as the anomaly feature can be delineated.
}
% conclusions heading (optional), leave it empty if necessary 
{}


\maketitle

\section{Introduction}\label{sec:one}

The microlensing signal of a planet generally appears as a short-term anomaly to the lensing 
light curve produced by the host of the planet \citep{Mao1991, Gould1992}. The planetary signal 
is produced when a source star approaches close to or passes over the lensing caustic induced 
by the planet. A planetary companion induces two sets of caustics, in which one tiny set forms 
near the host of the planetary system, and the other set forms in the region away from the host 
with a separation $\sim {\bf s}-1/{\bf s}$, where ${\bf s}$ denote the planet-host separation 
vector. The former and latter caustics are referred to as "central" and "planetary" caustics, 
respectively. Due to the caustic location, the planetary signal generated by the central caustic 
appears near the peak of a high-magnification event, while the signal produced by the planetary 
caustic can appear at any part of the lensing light curve.

In the early phase of the planetary microlensing experiments, the majority of microlensing 
planets were found from the observations of high-magnifications events, for example, 
OGLE-2005-BLG-071Lb \citep{Udalski2005}, OGLE-2005-BLG-169Lb \citep{Gould2006}, and 
MOA-2007-BLG-192Lb \citep{Bennett2008}. This was because the observational cadence of 
the microlensing surveys at that time was generally not high enough to detect short-term 
planetary signals, and thus planet searches were carried out in a survey+followup mode, in 
which survey groups mainly concentrated on event detections, and followup groups conducted 
high-cadence followup observations for a fraction of events found by the survey groups.  
In this observation mode, high-magnification events were important targets because the 
chance for the source stars of these events to pass the perturbation regions induced by 
the central caustics was very high and the time of the event peak could be predicted in 
advance for the preparation of followup observations \citep{Griest1998}.

With the launch of high-cadence microlensing surveys in the 2010s, not only the total 
detection rate of microlensing planets but also the fraction of planets detected via 
the perturbations induced by planetary caustics have greatly increased, for example, 
KMT-2016-BLG-1105Lb, KMT-2017-BLG-1194Lb, and OGLE-2017-BLG-1806 \citep{Zang2023}, 
KMT-2018-BLG-0173Lb \citep{Jung2022}, KMT-2021-BLG-0712Lb and KMT-2021-BLG-0909Lb 
\citep{Ryu2023}. The total planet detection rate has increased because all lensing 
events can be densely monitored regardless of lensing magnifications without the need of 
extra followup observations. The rate of planet detections via the 
signals generated by planetary caustics
has increased because all parts of lensing light curves can be monitored 
with moderately-high to very-high cadences.

The classic type of degeneracy in the interpretations of the signals induced by planetary 
caustics is the "inner-outer" degeneracy, which was predicted by \citet{Gaudi1997} even 
before the detection of the first microlensing planet OGLE 2003-BLG-235/MOA-2003-BLG-53Lb 
\citep{Bond2004}.  This degeneracy is intrinsic in the sense that it arises due to the 
intrinsic similarity between the magnification patterns on the near (to the primary lens) 
and far sides of the planetary caustic, and thus the planetary signal resulting from the 
source trajectory passing on near side of the caustic would be similar to the signal 
resulting from the trajectory passing on the far side.  The origins of this degeneracy 
have been further explored in \citet{Zhang2022}.

As the number of planets detected from actually observed signals generated by planetary 
increases, various types of unexpected degeneracies have been reported. From the analysis 
of the planetary lensing event OGLE-2017-BLG-0173, \citet{Hwang2018} reported a new type 
of degeneracy between the solution in which the source fully enveloped the caustic and 
the other solution in which the source enveloped one side of the caustic.  They also 
reported a degeneracy between the solution in which the light-curve perturbation was 
generated by a planetary caustic due to a "close" planet and the one in which it was 
generated by a "wide" planet.  Here the terms "close" and "wide" refer to the cases in 
which the projected planet-host separation is less and greater than the radius of the 
Einstein ring ($\thetae$), respectively.  From the analysis of the anomaly appearing in 
the planetary lensing event OGLE-2017-BLG-0373, \citet{Skowron2018} also reported multiple 
degeneracies that had not been known before.  We note that these new types of degeneracies 
were identified from the analyses of partially covered planetary signals, and thus it is 
not yet clear whether these degeneracies persist in the interpretations of continuously 
covered signals.

In this paper, we present the analysis of the lensing event KMT-2021-BLG-1150, for which a
short-term signal induced by a planetary caustic appears on the lensing light curve. Despite 
the short duration, the anomaly was densely and continuously covered including both the caustic
entrance and exit, and thus the event provides an important test bed that enables one to check
whether various types of recently-reported degeneracies persist even for well covered planetary
signals.

We present the analysis according to the following organization. In Sect.~\ref{sec:two}, we 
describe the observations of the lensing event and the data acquired from the observations. 
We mention the characteristics of the lensing event and the anomaly appearing in the lensing 
light curve. In Sect.~\ref{sec:three}, we depict the detailed procedure of the analysis and 
present the results found from the analysis. In Sect.~\ref{sec:four}, we explain the source 
characterization procedure conducted to estimate the angular Einstein radius and present the 
measured value of $\thetae$. In Sect.~\ref{sec:five}, we detail the Bayesian analysis conducted 
to estimate the physical parameters of the planetary system and present the estimated parameters. 
In Sect.~\ref{sec:six}, we present the results of the test conducted using mock data sets prepared 
to mimic those obtained with data gaps and under various observational cadences.  We summarize the 
results of the analysis and conclude in Sect.~\ref{sec:seven}.

% Figure 1 ------------------------------------------------------
\begin{figure}[t]
\includegraphics[width=\columnwidth]{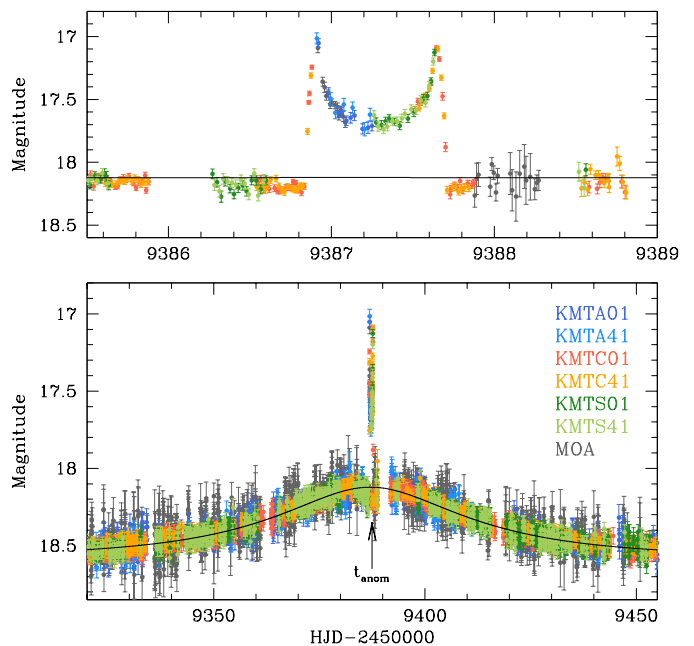}
\caption{
Light curve of KMT-2021-BLG-1150. The lower panel shows the whole view and the upper panel 
shows the zoom of the region around the anomaly.  The curve drawn over the data point is a 
1L1S model obtained by fitting the light curve excluding the data points around the anomaly.
}
\label{fig:one}
\end{figure}
% --------------------------------------------------------------

\section{Observations and data}\label{sec:two}

The source of the microlensing event KMT-2021-BLG-1150 lies toward the Galactic bulge field 
at the equatorial coordinates $({\rm RA}, {\rm DEC})_{\rm J2000} = ($17:55:09.63, 
-31:18:47.99), which correspond to the Galactic coordinates $(l, b) = (-0^\circ\hskip-2pt 
.986, -2^\circ\hskip-2pt .989)$.  The extinction toward the field is $A_I = 1.50$, and the 
baseline $I$-band magnitude of the source is $I_{\rm base} = 18.86$, as derived from the 
calibrated OGLE-III catalog \citep{Szymanski2011}.  The magnification of the source flux 
induced by lensing was first found by the EventFinder system \citep{Kim2018} of the KMTNet 
group \citep{Kim2016} on 2021 June 4, which corresponds to the abridged heliocentric Julian 
date ${\rm HJD}^\prime  = {\rm HJD}- 2450000 = 9369.56$.  On 2021 June 21, the event was 
independently found by the MOA group \citep{Bond2001}, who labeled the event as MOA-2021-BLG-197. 
Hereafter we refer to the event as KMT-2021-BLG-1150 following the convention of using the event 
ID reference of the first discovery group. The source lies in the KMTNet prime fields BLG01 
and BLG41, toward which observations were conducted with a 0.5~hr cadence for each field and 
0.25~hr in combination.  The MOA observations were carried out at a similar cadence.

Observations by the KMTNet group were done utilizing the three identical 1.6~m telescopes that
are globally distributed in the Southern Hemisphere at the Cerro Tololo Inter-American Observatory
in Chile (KMTC), the South African Astronomical Observatory in South Africa (KMTS), and the
Siding Spring Observatory in Australia (KMTA). The MOA group used the 1.8~m telescope located
at the Mt. John Observatory in New Zealand. Images were acquired mainly in the $I$ band for
the KMTNet survey and in the customized MOA-$R$ band for the MOA survey. For both surveys,
subsets of images were taken in the $V$ band for the purpose of measuring the color of the
source.

% Figure 2 ------------------------------------------------------
\begin{figure}[t]
\includegraphics[width=\columnwidth]{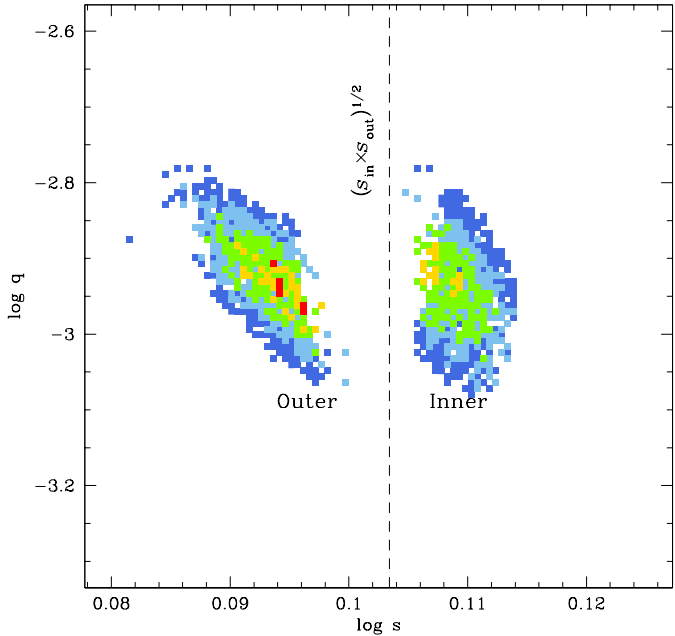}
\caption{
Map of $\Delta\chi^2$ in the space of the binary-lens parameters $s$ and $q$. The vertical 
dashed line indicates the geometric mean of $s_{\rm in}$ and $s_{\rm out}$, which represent 
the binary separations of the inner and outer solutions, respectively.  The color coding is 
set to represent points with 
$\leq 1n\sigma$ (red),
$\leq 2n\sigma$ (yellow),
$\leq 3n\sigma$ (green),
$\leq 4n\sigma$ (cyan) and,
$\leq 5n\sigma$ (blue), where $n=3$.
}
\label{fig:two}
\end{figure}
% --------------------------------------------------------------

Figure~\ref{fig:one} shows the lensing light curve of KMT-2021-BLG-1150 constructed by 
combining the data from the KMTNet and MOA surveys. It displays a short-term anomaly 
appearing near the peak of the event. The curve drawn over the data points is the 
single-lens single-source (1L1S) model obtained by fitting the data excluding those around 
the anomaly. From the zoom-in view of the anomaly region presented in the top panel, it is 
found that the anomaly exhibits two caustic-crossing features at $t_{\rm c,1}\sim 9386.9$ 
(caustic entrance) and $t_{\rm c,2}\sim 9387.7$ (caustic exit) with a time gap $\Delta 
t_{\rm c} = t_{\rm c,2}-t_{\rm c,1}\sim  0.8$~day between the caustic features. Besides 
these caustic features, the anomaly exhibits negative deviations of $\sim 0.1$~mag level 
from the 1L1S model just before the caustic entrance and after the caustic exit. We note 
that all these anomaly features were densely resolved by the data from the combined 
high-cadence observations of the survey groups.

In the analysis of the event, we used the data prepared by reducing the images and conducting
photometry of the source with the use of the pipelines of the individual groups. The pySIS 
pipeline of the KMTNet group was developed by \citet{Albrow2009}, and that of the MOA survey 
was developed by \citet{Bond2001}. Both pipelines commonly applied the difference image technique 
\citep{Tomaney1996, Alard1998} that was developed for the optimal photometry of stars lying 
in very dense fields. Following the \citet{Yee2012} routine, we readjusted the error bars 
estimated by the pipelines so that the scatter of data is consistent with the error bars and 
the $\chi^2$ value per degree of freedom (dof) for each data becomes unity.

% Table 1 ------------------------------------------------
\begin{table}[t]
\small
%\centering
\caption{Model parameters\label{table:one}}
\begin{tabular*}{\columnwidth}{@{\extracolsep{\fill}}lcccc}
\hline\hline
\multicolumn{1}{c}{Parameter}    &
\multicolumn{1}{c}{Inner}        &
\multicolumn{1}{c}{Outer}        \\
\hline
$\chi^2$/dof            &   $8139.7/8152        $  &  $8133.4/8152        $   \\
$t_0$ (HJD$^\prime$)    &   $9387.077 \pm 0.055 $  &  $9387.086 \pm 0.055 $   \\
$u_0$                   &   $0.502 \pm 0.010    $  &  $0.464 \pm 0.010    $   \\
$\te$ (days)            &   $41.55 \pm 0.58     $  &  $43.75 \pm 0.65     $   \\
$s  $                   &   $1.297 \pm 0.005    $  &  $1.242 \pm 0.007    $    \\
$q  $ ($10^{-3}$)       &   $1.10 \pm 0.05      $  &  $1.15 \pm 0.05      $    \\
$\alpha$ (rad)          &   $4.703 \pm 0.003    $  &  $4.703 \pm 0.003    $    \\
$\rho$ ($10^{-3}$)      &   $0.956 \pm 0.015    $  &  $0.899 \pm 0.016    $    \\
\hline
\end{tabular*}
\tablefoot{ ${\rm HJD}^\prime = {\rm HJD}- 2450000$.  }
\end{table}
% --------------------------------------------------------

\section{Interpretation of the anomaly}\label{sec:three}

The anomaly in the lensing light curve of KMT-2021-BLG-1150 is characterized by a near-peak 
caustic feature. Such an anomaly feature is known to be produced by 3 major channels. The 
first is the "high-magnification" channel, in which an anomaly near the peak is produced 
when a source approaches close to the central caustic induced either by a planetary companion 
lying around the Einstein ring ($s\sim 1$) or a binary companion with a very small ($s\ll 1$) or 
large ($s\gg 1$) separation from the primary \citep{Han2009}. The anomaly produced via this 
channel arises for a high-magnification event because the central caustic is very tiny 
\citep{Chung2005}, and thus the source should approach very close to the primary for the 
production of a near-peak anomaly. The second is the "right-angle" channel, in which a 
near-peak anomaly is generated when a source passes the perturbation region around the 
caustic at an approximately right angle with respect to the planet-host axis, for example, 
KMT-2021-BLG-0748 \citep{Ryu2022} and KMT-2021-BLG-1554 \citep{Han2022}. In this case, the 
anomaly can arise regardless of the peak magnification because the caustic can appear at any 
part around the Einstein ring. The third is the "off-axis" channel, in which a near-peak 
anomaly is produced by the source passage through the anomaly region around the caustic lying 
away from the binary axis, for example, OGLE-2016-BLG-1266 \citep{Albrow2018} and the off-axis 
solution of KMT-2018-BLG-1497 \citep{Jung2022}. Among these channels, the anomaly in 
KMT-2021-BLG-1150 is unlikely to be produced by the high-magnification channel because it 
appears when the magnification $A_{\rm anom}\sim 2.4$ is not very high.

% Figure 3 ------------------------------------------------------
\begin{figure*}[t]
\centering
\includegraphics[width=13.5cm]{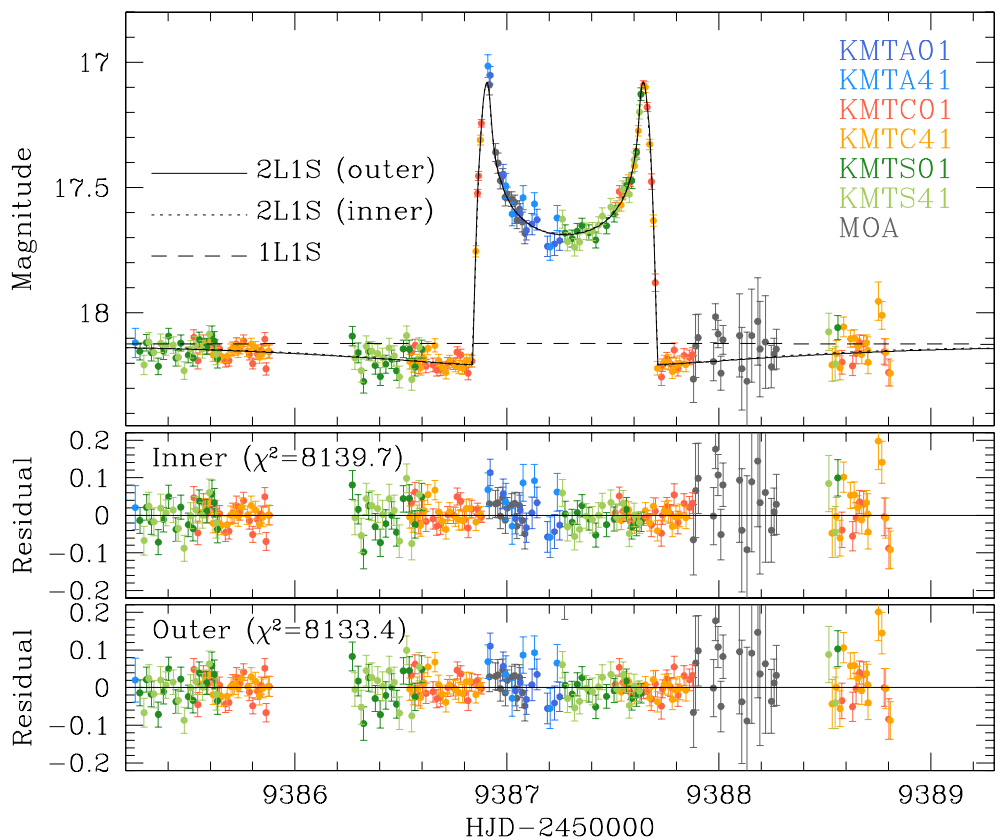}
\caption{
Enlarged view around the region of the anomaly. The dashed, dotted, and solid curves represent 
the models of the 1L1S, inner 2L1S, and outer 2L1S solutions, respectively.  We note that the 
model curves of the inner and outer 2L1S solutions are so similar that it is difficult to 
resolve them with the given line width.
}
\label{fig:three}
\end{figure*}
% --------------------------------------------------------------

For the interpretation of the anomaly, we conducted modeling of the observed lensing light 
curve. In the modeling, we searched for a lensing solution representing a set of the lensing 
parameters that best described the observed anomaly. Considering that caustic features are 
produced by the multiplicity of a lens, we conducted a binary-lens single-source (2L1S) 
modeling. The lensing parameters of the 2L1S model include $(t_0, u_0, \te, s, q, \alpha, 
\rho)$. The first three of these parameters describe the source approach to the lens, and 
they represent the time of the closest lens-source approach, the separation at that time, 
and the event time scale, respectively. The next three parameters describe the binarity of 
the lens, and the individual parameters denote the projected separation (scaled to $\thetae$) 
and mass ratio between the lens components, and the source trajectory angle defined as the 
angle between the source trajectory and the binary axis. The last parameter denotes the ratio
of the angular source radius $\theta_*$ to the Einstein radius, that is, $\rho= \theta_*/\thetae$ 
(normalized source radius), and this parameters is needed to describe the deformation of the 
caustic-crossing features by finite-source effects.

Considering that various types of degeneracies had been reported from the recent analyses 
of microlensing planets detected through the signals generated by planetary caustics, we 
thoroughly investigated the parameter space to check all possible degenerate solutions. 
Figure~\ref{fig:two} shows the $\Delta\chi^2$ map constructed by conducting dense grid 
searches for the binary-lens parameters $s$ and $q$.  The map shows two distinct local 
solutions resulting from the inner-outer degeneracy with planetary parameters $(s, q)_{\rm in}
\sim (1.297, 1.10\times 10^{-3})$ for the inner solution and $(s, q)_{\rm out}\sim (1.242, 
1.15\times 10^{-3})$ for the outer solution. The degeneracy between the two solutions is 
fairly severe, although the outer solution is slightly preferred over the inner solution
 by $\Delta\chi^2=6.3$.

% Figure 4 ------------------------------------------------------
\begin{figure}[t]
\includegraphics[width=\columnwidth]{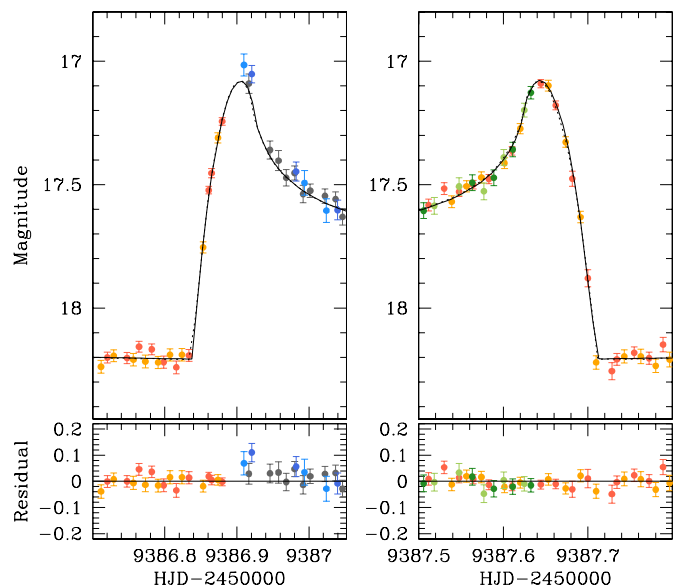}
\caption{
Zoom-in views around the caustic entrance (left panel) and exit (right panel). The solid 
curve is the model curve of the outer 2L1S solution.
}
\label{fig:four}
\end{figure}
% -------------------------------------------------------------

We list the full lensing parameters of the inner and outer 2L1S solutions in Table~\ref{table:one} 
together with the $\chi^2$ values of the fits and degrees of freedom.  The two degenerate solutions 
result in similar mass ratios of $q\sim 1.1 \times 10^{-3}$, and the low value of $q$ indicates 
that the companion to the lens is a planetary mass object.  The planet separations of the inner 
and outer solutions are slightly different from each other, and the separation of the inner 
solution $s_{\rm in}=1.297 \pm 0.005$ is slightly larger than the separation of the outer 
solution of $s_{\rm out}=1.242 \pm 0.007$. Besides these two solutions, we found no other 
degenerate solutions. For a double check, we investigated a degenerate solution resulting 
from the off-axis channel by restricting the source trajectory to pass the off-axis cusps, 
and found that the best-fit off-axis solution resulted in a model that yielded a poorer fit 
than the outer solution by $\Delta\chi^2=2320$.

In Figure~\ref{fig:three}, we present the model curves of the inner (dotted curve) and outer 
(solid curve) solutions around the region of the anomaly. It shows that both models well 
describe the anomaly including the positive deviations caused by caustic crossings and 
the negative deviations before the caustic entrance and after the caustic exit. As shown in 
the zoom-in views around the caustic entrance and exit presented in Figure~\ref{fig:four}, 
both caustic crossings were densely resolved despite the short duration of each caustic 
crossing of $\Delta t_{\rm c}\sim 2\rho \te/\sin \psi\sim 2$~hr, where $\psi \sim 70^\circ$ 
represents the incidence angle of the source to the fold of the caustic. With the well 
resolved caustic crossings, the normalized source radius was precisely measured.

\begin{figure}[t]
\includegraphics[width=\columnwidth]{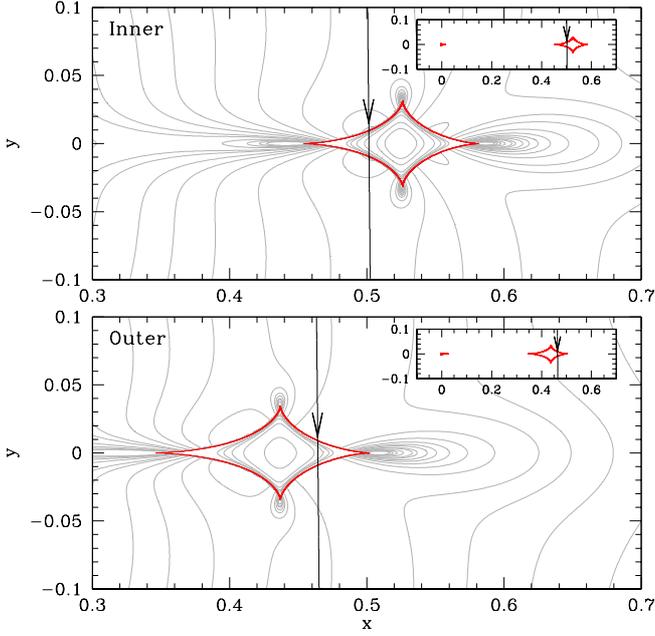}
\caption{
Lens-system configurations of the inner (upper panel) and outer (lower panel) solutions. 
The inset in each panel shows the whole view of the caustic (red figures), and the main 
panel shows the zoom-in view around the planetary caustic. The arrowed line represents 
the source trajectory, and the grey curves encompassing the caustic are equi-magnification 
contours.  
}
\label{fig:five}
\end{figure}

The lens-system configurations of the inner and outer 2L1S solutions are presented in 
Figure~\ref{fig:five}, which shows the trajectory of the source with respect to the lens 
caustic. It shows that the anomaly was produced by the crossings of the source over the 
planetary caustic formed by a planetary companion, and the source crossed the near and 
far side according to the inner and outer solutions, respectively. The source crossed the 
planet-host axis at a very nearly right angle, 89.4 deg, and this explains the near-peak 
location of the anomaly. The lens-source separation (scaled to $\thetae$) at the time of the 
anomaly, $u_{\rm anom}\sim 0.5$, is not small, and this explains the low lensing magnification 
at the time of the anomaly.

% Figure 5 ------------------------------------------------------
% --------------------------------------------------------------

We found that the pair of the planet separations of the inner and outer solutions, $(s_{\rm in}, 
s_{\rm out})$, well obeyed the analytic relation introduced by \citet{Hwang2022}. The relation 
is expressed as
\begin{equation}
s^\dagger = \sqrt{s_{\rm in}\times s_{\rm out}} =
{\sqrt{u_{\rm anom}^2+4}+u_{\rm anom}  \over 2},
\label{eq1}
\end{equation}
where $u^2_{\rm anom}=\tau^2_{\rm anom}+u_0^2$, 
$\tau_{\rm anom}=(t_{\rm anom}-t_0)/\te$, and $t_{\rm anom}$ represents the time of the anomaly. 
We note that the second term in Equation~(\ref{eq1}) was originally expressed as the arithmetic 
mean of $s_{\rm in}$ and $s_{\rm out}$, that is, $(s_{\rm in}+s_{\rm out})/2$, and \citet{Gould2022} 
found that the geometric mean, this is, $(s_{\rm out}\times s_{\rm in})^{1/2}$, proved to be a 
better approximation than the arithmetic mean.  The $s^\dagger$ value estimated from $s_{\rm in}$ 
and $s_{\rm out}$ is $s^\dagger=(s_{\rm out}\times s_{\rm in})^{1/2}= 1.269$.  From the lensing 
parameters $(t_0, t_{\rm anom}, \te)\sim (9387.1, 9387.3, 43.5)$, it is estimated that 
$\tau_{\rm anom} = (t_{\rm anom}-t_0)/\te = 0.005$, $u_{\rm anom} = (\tau_{\rm anom}^2 +
u_0^2)^{1/2} = 0.480$, and $s^\dagger=[(u_{\rm anom}^2+4)^{1/2}+u_{\rm anom}]/2 = 1.269$. 
Therefore, the $s^\dagger$ values estimated from the planet separations $(s_{\rm in}, 
s_{\rm out})$ and from the planetary parameters $(t_0, t_{\rm anom}, u_0, \te)$ match very well 
down to 3 digits after the decimal point, proving the validity of the analytic relation. In 
Figure~\ref{fig:two}, we mark the geometric mean of $s_{\rm in}$ and $s_{\rm out}$ as a dashed 
vertical line.

We checked whether the microlens parallax $\pivec_{\rm E}$ could be measured by conducting an 
additional modeling including the two microlens-parallax parameters $(\pien, \piee)$, which 
denote the north and east components of the microlens-parallax vector $\pivec_{\rm E}=
(\pivec_{\rm rel}/ \thetae)(\muvec/\mu)$, respectively \citep{Gould1992a, Gould2000, Gould2004}.  
Here $\muvec$ represents the relative lens-source proper motion vector, $\pi_{\rm rel}=\pi_{\rm L}
-\pi_{\rm S} ={\rm AU}(1/\dl-1/\ds)$ is the relative lens-source parallax, and $(\dl, \ds)$ denote 
the distances to the lens and source, respectively.  We found that it was difficult to measure the 
parallax parameters because the lensing magnification of the event was low, and thus the light 
curve was not susceptible to the subtle variation induced by higher-order effects.

% Figure 6 ------------------------------------------------------
\begin{figure}[t]
\includegraphics[width=\columnwidth]{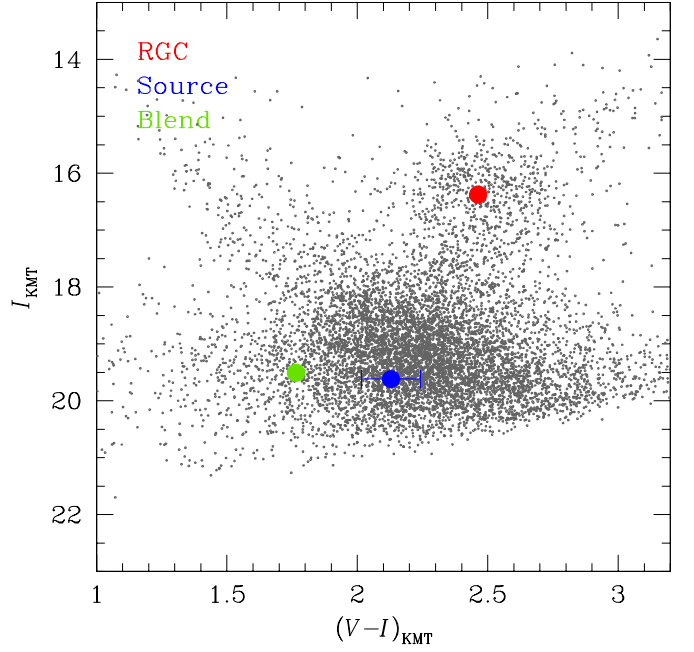}
\caption{
Locations of the source, blend, and red giant clump (RGC) centroid in the instrumental 
color-magnitude diagram.
}
\label{fig:six}
\end{figure}
% --------------------------------------------------------------

\section{Source star and Einstein radius}\label{sec:four}

In this section, we characterize the source star of the event and estimate the angular Einstein 
radius. We estimated $\thetae$ from the relation
\begin{equation}
\thetae = {\theta_* \over \rho},
\label{eq2}
\end{equation}
where the angular source radius $\theta_*$ was deduced from the source type, and the normalized 
source radius $\rho$ was measured from modeling the light curve. For the specification of the 
source type, we estimated the reddening- and extinction-corrected (de-reddened) color and 
magnitude of the source, $(V-I, I)_{{\rm S},0}$.

We used the \citet{Yoo2004} method for the estimation of $(V-I, I)_{{\rm S},0}$.  Following the 
routine procedure of the method, we first measured the instrumental color and magnitude of the 
source, placed the source in the instrumental color-magnitude diagram (CMD), and then 
calibrated the source color and magnitude using the centroid of the red giant clump (RGC) in the 
CMD as a reference.  The RGC centroid can serve as a reference for calibration because of its 
known de-reddened color and magnitude.

In Figure~\ref{fig:six}, we mark the source position with respect to the RGC centroid in the 
instrumental CMD constructed using the pyDIA photometry of stars lying around the source. The 
$I$- and $V$-band magnitudes of the source were measured by regressing the pyDIA photometry 
data of the individual passbands with respect to the model light curve. The measured colors 
and magnitudes of the source and RGC centroid are $(V-I, I)_{\rm S} = (2.129\pm 0.113, 19.619
\pm 0.008)$ and $(V-I, I)_{\rm RGC} = (2.463\pm 0.020, 16.377\pm 0.040)$, respectively.  We 
note that the magnitude of the instrumental CMD is approximately scaled, and thus the extinction 
toward field measured from $I_{\rm RGC}-I_{\rm RGC,0}=1.87$ does not match the value $A_I=1.50$ 
that is estimated from the calibrated magnitude.  The measurement uncertainty of $V-I$ is fairly 
big because the source color was measured mainly on a single $V$-band point inside the caustic 
trough taken at ${\rm HJD}^\prime=9387.584$.  With the known de-reddened values of the RGC 
centroid, $(V-I, I)_{\rm RGC,0} = (1.060, 14.502)$ \citep{Bensby2013, Nataf2013}, then the 
de-reddened source color and magnitude were estimated as 
\begin{equation}
\eqalign{
(V-I, I)_{\rm 0} =  & \ (V-I, I)_{{\rm RGC},0} + \Delta(V-I, I) \cr
                 =  & \ (0.726\pm 0.120, 17.744\pm 0.021),      \cr
}
\label{eq3}
\end{equation}
where $\Delta(V-I, I) = (V-I, I)_{\rm S} - (V-I, I)_{\rm RGC}$ represents the offset of the 
source from the RGC centroid.  According to the estimated color and magnitude, the source is 
a G-type main-sequence star.  We note that the error of the dereddened $I$-band magnitude, 
that is, $\sigma(I_0)=0.021$~mag, is based purely on the fractional error of the source flux 
measured from the modeling.  \citet{Gould2014} pointed out that the $\theta_*$ measurement is 
subject to additional errors originating from the uncertain dereddened RGC color of 
\citet{Bensby2013} and the uncertain position of the RGC, and these two errors combined yield 
a 7\% error in the estimation of $\theta_*$. In our estimation of $\theta_*$, we add this 
fractional error.

Once the source type was specified, we then estimated the source radius.  For the $\thetae$ 
estimation, we use the \citet{Adams2018} relation of 
\begin{equation}
\theta_* = 0.5\times 10^{[-0.2 I_0 + c_0 + c_1 (V-I)_0 + 3]} ~\mu{\rm as}, 
\label{eq4}
\end{equation}
where the values of the coefficients are
$c_0 = 0.542\pm 0.007$ and $c_1 = 0.378 \pm 0.011$.
The measured source radius is
\begin{equation}
\theta_* = 0.93 \pm  0.13~\mu{\rm as},
\label{eq5}
\end{equation}
and this yields the angular Einstein radius 
\begin{equation}
\thetae = 1.03 \pm 0.14~{\rm mas}.
\label{eq6}
\end{equation}
With the measured value of $\thetae$ together with the event time scale $\te$, we estimated the 
relative lens-source proper motion as
\begin{equation}
\mu = {\thetae\over \te} = 8.59 \pm 1.20~{\rm mas}/{\rm yr}.
\label{eq7}
\end{equation}

Also marked in Figure~\ref{fig:six} is the position of the blend (green filled dot), which lies 
at $(V-I, I)_{\rm b} = (1.76, 19.50)$. We checked the possibility that the lens would be the main 
source of the blended flux by measuring the astrometric offset between the centroid of the source 
image obtained before the lensing magnification and the centroid of the image obtained at the peak 
of the lensing magnification. The measured offset is $\delta\theta = (275.95 \pm 129.60)$~mas. 
Considering that the offset is greater than the measurement uncertainty by a factor 2.1, it is 
unlikely that the main origin of the blended flux is the lens.

% Figure 7 ------------------------------------------------------
\begin{figure}[t]
\includegraphics[width=\columnwidth]{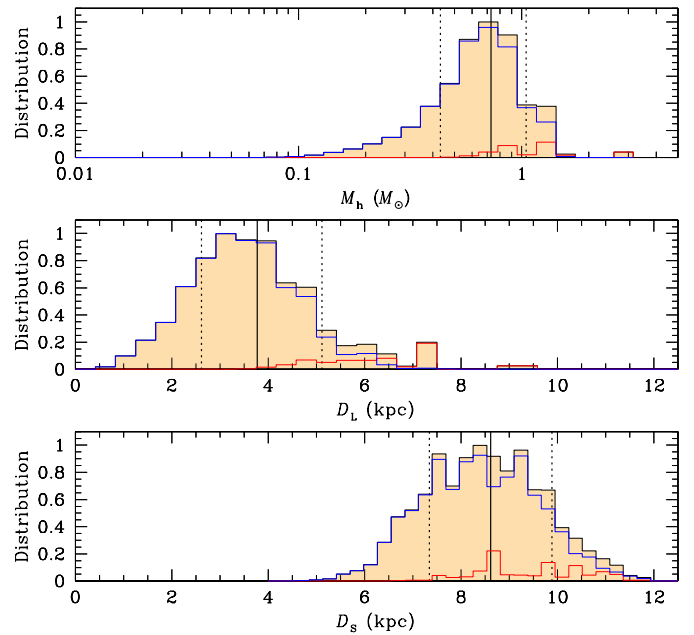}
\caption{
Bayesian posteriors of the primary lens mass and distances to the lens and source.  In each panel, 
the solid vertical line represent the median value, and the two dotted lines represent the 16\% 
and 84\% of the distribution.
}
\label{fig:seven}
\end{figure}
% --------------------------------------------------------------

\section{Physical lens parameters}\label{sec:five}

In this section, we estimate the physical parameters of the mass and distance to the lens. For 
the unique determinations of these parameters, it is required to simultaneously measure the three 
lensing observables $(\te, \thetae, \pie)$, from which the parameters are determined by the
relations 
\begin{equation}
M={\thetae \over \kappa\pie};\qquad
\dl = {{\rm AU}  \over \pie\thetae + \pi_{\rm S}},
\label{eq8}
\end{equation}
where $\kappa = 4G/(c^2{\rm AU})\simeq 8.14~{\rm mas}/M_\odot$, and $\pi_{\rm S}={\rm AU}/\ds$ 
denotes the parallax of the source \citep{Gould2000}. For KMT-2021-BLG-1150, the lensing observables 
$\te$ and $\thetae$ were securely measured, but the other parameter $\pie$ could not be measured, 
making it difficult to uniquely determine $M$ and $\dl$ using the analytic relations in 
Equation~(\ref{eq8}). We, therefore, estimated the physical parameters of the planetary system by 
conducting a Bayesian analysis based on the measured observables and using a Galactic model.  In 
the analysis, we additionally imposed the lens-brightness constraint given by the fact that the 
lens cannot be brighter than the blend. It turned out that this constraint had little effect on 
the Bayesian posteriors.

% Table 2 ------------------------------------------------
\begin{table}[t]
\small
%\centering
\caption{Physical lens parameters\label{table:two}}
\begin{tabular*}{\columnwidth}{@{\extracolsep{\fill}}lcccc}
\hline\hline
\multicolumn{1}{c}{Parameter}    &
\multicolumn{1}{c}{Value}        \\                                                    [0.8ex]
\hline 
$M_{\rm h}$ ($M_\odot$)    &  $ 0.73^{+0.32}_{-0.30}$    \\           [0.8ex]
$M_{\rm p}$ ($M_{\rm J}$)  &  $ 0.88^{+0.38}_{-0.36}$    \\           [0.8ex]
$\dl$ (kpc)                &  $ 3.8^{+1.3}_{-1.2}   $    \\           [0.8ex]
$a_\perp$ (AU)             &  $ 4.5^{+1.6}_{-1.4}   $    \\    [0.8ex]
\hline
\end{tabular*}
\end{table}
% --------------------------------------------------------

% Table 3 ------------------------------------------------
\begin{table*}[t]
\small
%\centering
\caption{Variation of planet parameters depending on data\label{table:three}}
\begin{tabular}{l|ll|ll}
\hline\hline
\multicolumn{1}{c|}{Cadence}              &
\multicolumn{2}{c|}{Inner}                &
\multicolumn{2}{c}{Outer}                \\
\multicolumn{1}{c|}{}                     &
\multicolumn{1}{c}{$s$}                  &
\multicolumn{1}{c|}{$q$ ($10^{-3}$)}      &
\multicolumn{1}{c}{$s$}                  &
\multicolumn{1}{c}{$q$ ($10^{-3}$)}      \\
\hline
 0.25 hr &   $1.297 \pm 0.005$  &  $1.10 \pm 0.05 $   &   $1.242 \pm 0.007$  &  $1.15 \pm 0.05$   \\           
 1.0 hr  &   $1.280 \pm 0.008$  &  $1.29 \pm 0.08 $   &   $1.227 \pm 0.010$  &  $1.26 \pm 0.06$   \\           
 2.5 hr  &   $1.297 \pm 0.009$  &  $1.30 \pm 0.09 $   &   $1.273 \pm 0.017$  &  $1.00 \pm 0.15$   \\           
 5.0 hr  &   $1.297 \pm 0.013$  &  $1.35 \pm 0.12 $   &   $1.278 \pm 0.024$  &  $1.08 \pm 0.18$   \\           
\hline
\end{tabular}
%\tablefoot{ }
\end{table*}
% --------------------------------------------------------

In the Bayesian analysis, we started by generating a large number ($2\times 10^6$) of artificial 
lensing events from the Monte Carlo simulation conducted using a Galactic model and a mass 
function model of lens objects. The Galactic model defines the physical and dynamical distributions 
of Galactic objects. In the simulation, we adopted the Galactic and mass function models described 
in \citet{Jung2021}. For each simulated event with $(M_i, D_{{\rm L},i}, \mu_i)$, we computed the 
lensing observables using the relations $t_{{\rm E},i} = \theta_{{\rm E},i}/\mu_i$ and 
$\theta_{{\rm E},i} = (\kappa M_i \pi_{{\rm rel},i})^{1/2}$, where $\pi_{{\rm rel},i}={\rm AU}
(1/D_{{\rm L},i}-1/D_{{\rm S},i})$.  Based on the simulated events, posterior distributions of $M$ 
and $\dl$ were constructed by imposing a weight $w_i=\exp(-\chi^2/2)$ to each artificial event. 
Here the $\chi^2$ value is computed by $\chi^2 =(t_{{\rm E},i}-\te)^2/\sigma(\te)^2+ 
(\theta_{{\rm E},i}-\thetae)^2/\sigma(\thetae)^2$, where $(\te, \thetae)$ represent the measured 
observables and $[\sigma(\te), \sigma(\thetae)]$ denote their measurement uncertainties.

Figure~\ref{fig:seven} shows the Bayesian posteriors of the mass of the planet host and distances 
to the lens and source.  The physical parameters of the planetary system, including the masses of 
the host ($M_{\rm h}$) and planet ($M_{\rm p}$), distance, and projected physical separation 
($a_\perp$) between the planet and host, are listed in Table~\ref{table:two}. According to the 
Bayesian estimation, the lens is a planetary system consisting of a giant planet with a mass 
$M_{\rm p}= 0.88^{+0.38}_{-0.36}~M_{\rm J}$ and its host with a mass 
$M_{\rm h}=0.73^{+0.32}_{-0.30}~M_\odot$ lying toward the Galactic center at a distance 
$\dl =3.8^{+1.3}_{-1.2}$~kpc.  Here we adopted the median values of the posterior distributions 
as representative parameters and the uncertainties were estimated as the 16\% and 84\% of the 
distributions.  The projected separations estimated from the inner and outer solutions are 
$a_{\perp,{\rm in}}\sim 4.4$~AU and $a_{\perp,{\rm out}}\sim 4.6$~AU, respectively, and this 
indicates that the planet lies beyond the snow line, $a_{\rm sl}\sim 2.7(M_{\rm h} /M_\odot){\rm AU} 
\sim 2.0$~AU, of the planetary system regardless of the solutions.  Although slightly different, 
the difference between $a_{\perp, {\rm in}}$ and $a_{\perp, {\rm out}}$ is within the uncertainty 
of each, and thus we present the mean value in Table~\ref{table:two}.  We found that the probabilities 
for the lens to be in the disk and bulge are 93\% and 7\%, respectively, and thus the host of the 
planetary system is very likely to be a disk star.

\section{Discussion}\label{sec:six}

Degeneracies in the interpretation of a planetary signal may be caused due to either the sparse
coverage of the signal or data gaps in the signal. Data gaps can arise when observations cannot 
be done due to bad weather or when there are gaps between the end of the night at one telescope
site and the beginning of the night at another site. In this section, we investigate how the
observational cadence and data gaps in microlensing data affect the interpretation of planetary
signals by conducting analyses based on mock data sets prepared using those of KMT-2021-BLG-1150 
to mimic events with sparse and incomplete data coverage.

\begin{figure}[t]
\includegraphics[width=\columnwidth]{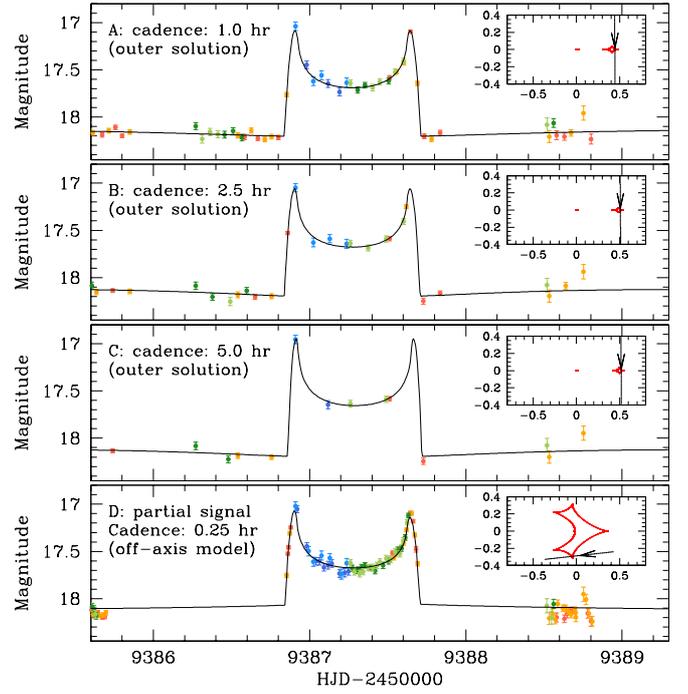}
\caption{
Four sets of mock data. The data sets in the top three panels mimic those obtained with a 1.0~hr, 
a 2.5~hr, and a 5.0~hr cadence, and the data set in the bottom panel simulate partially covered data. 
The inset in each panel shows the lens-system configuration of the solution found from modeling, and 
the solid curve drawn on the data points is the model curve.
}
\label{fig:eight}
\end{figure}

Figure~\ref{fig:eight} shows the anomaly regions of 4 data sets, in which the first 3 sets 
are generated to mimic data obtained with a 1.0~hr, a 2.5~hr, and a 5.0~hr cadence, and the last 
one was created to mimic a data set that misses the negative deviations that occur just before 
the caustic entrance and just after the caustic exit.  The three tested observational cadences 
correspond to those of the KMTNet sub-prime fields, and the assumed 0.25~hr cadence of the last 
data set corresponds to the cadence of the KMTNet prime field. We note that the anomaly feature 
is still delineated in the 5-hr-cadence data set.

% Figure 8 ------------------------------------------------------
% --------------------------------------------------------------

In Table~\ref{table:three}, we list the planet parameters $(s, q)$ obtained by conducting modeling 
using the individual mock data sets. It shows that the inner and outer solutions are identified 
and the variation in the planet parameters among the solutions is not very big, although the 
measurement uncertainties of the lensing parameters increase as the observational cadence increases.  
The result of the simulation indicates that the observational cadence does not pose a serious 
degeneracy problem as long as the anomaly feature is delineated. In the insets of top three panels 
of Figure~~\ref{fig:eight}, we present the lens-system configurations of the outer solutions 
corresponding to the individual data sets.  From the analysis of the partially-covered data set, 
on the other hand, we were able to identify an off-axis solution, which was decisively rejected 
in the analysis conducted using the full data set.  The planet parameters of the off-axis solution 
are $(s, q)\sim (0.961 \pm 0.001, 0.047 \pm 0.001)$, which are substantially different from those 
of the correct solution. The lens system configuration of the off-axis solution is shown in the 
inset of the bottom panel in Figure~\ref{fig:eight}.  Although the fit of the off-axis solution 
is still substantially worse than the outer solution by $\Delta\chi^2\sim 1054$, one finds that 
the model curve of the off-axis solution appears to approximately describe the caustic features.  
The identification of the degenerate off-axis solution illustrates that gaps in data can cause 
degeneracy problems, and this conclusion is further supported by the fact that the recently 
reported various types of degeneracies were identified from the analyses of partially-covered 
data sets.

Although the inner and outer solutions are distinct, the effect of the degeneracy on the planet 
parameters is relatively not severe. This is because the planet parameters of the pair of the 
inner and outer solutions are similar to each other except for the planetary separation $s$, 
and even for $s$, the difference between the separations estimated by the two solutions is minor. 
In the case of KMT-2021-BLG-1150, this difference is $\Delta s=|s_{\rm in}-s_{\rm out}|=0.055$. 
The difference becomes smaller for lower-mass planets because the size of the planetary caustic 
decreases as the planet/host mass ratio becomes smaller \citep{Han2006}. By contrast, the variations 
in the planet parameters among the lensing solutions resulting from other degeneracy types can be 
substantial as illustrated by the above off-axis solution and by the recently-reported events with 
various types of degeneracies found from the analyses of partially covered planetary signals.

We note that the above test has been conducted on an event, that is,  KMT-2021-BLG-1150, with 
a relatively large mass ratio, $q\sim 10^{-3}$. The fact that the event parameters are recovered 
even when the cadence is greatly reduced supports the original decision of the KMTNet group to 
survey the majority of the bulge at a 2.5 hour cadence. By contrast, the 10-times higher-cadence 
observations of the prime fields are designed to capture $q\sim 10^{-5}$ planets, which are 100 
times less massive and so have planetary-caustic anomalies that evolve $100^{1/2}=10$ times faster. 
What is surprising, however, is that even modest gaps in high cadence observations can lead to 
major degeneracies, even when most of the caustic structure is densely covered.

\section{Summary and conclusion}\label{sec:seven}

We analyzed the microlensing event KMT-2021-BLG-1150, for which a densely and continuously resolved 
short-term anomaly appeared in the lensing light curve.  From the thorough investigation of the 
parameter space, we identified a pair of planetary solutions resulting from the well-known inner-outer 
degeneracy, and found that interpreting the anomaly was not subject to any degeneracy other than the 
inner-outer degeneracy.  The measured planet parameters are $(s, q)_{\rm in}\sim (1.297, 1.10\times 
10^{-3})$ for the inner solution and $(s, q)_{\rm out}\sim (1.242, 1.15\times 10^{-3})$ for the outer 
solution.  We found that the pair of the planet separations of the inner and outer solutions well 
obeyed the analytic relation introduced by \citet{Hwang2022}.  According to the physical parameters 
estimated from a Bayesian analysis, it was found that the lens is a planetary system consisting of 
a planet with a mass $M_{\rm p}=0.88^{+0.38}_{-0.36}~M_{\rm J}$ and its host with a mass $M_{\rm h} 
=0.73^{+0.32}_{-0.30}~M_\odot$ lying toward the Galactic center at a distance 
$\dl = 3.8^{+1.3}_{-1.2}$~kpc.  From the analyses conducted using mock data sets prepared to mimic 
those obtained with data gaps and under various observational cadences, it was found that gaps in 
data could result in various degenerate solutions, while the observational cadence would not pose 
a serious degeneracy problem as long as the anomaly feature were be delineated.

% --------------------------------------------------------------
\begin{acknowledgements}
Work by C.H. was supported by the grants of National Research Foundation of Korea 
(2020R1A4A2002885 and 2019R1A2C2085965).
% KMTNet
This research has made use of the KMTNet system operated by the Korea Astronomy and Space Science 
Institute (KASI) at three host sites of CTIO in Chile, SAAO in South Africa, and SSO in Australia. 
Data transfer from the host site to KASI was supported by the Korea Research Environment Open NETwork 
(KREONET).
This research was supported by the Korea Astronomy and Space Science Institute under the R\&D
program (Project No. 2023-1-832-03) supervised by the Ministry of Science and ICT.
% MOA
The MOA project is supported by JSPS KAKENHI
Grant Number JSPS24253004, JSPS26247023, JSPS23340064, JSPS15H00781,
JP16H06287, and JP17H02871.
%Yee
J.C.Y., I.G.S., and S.J.C. acknowledge support from NSF Grant No. AST-2108414. 
%Yossi Shvartzvald
Y.S.  acknowledges support from NSF Grant No. 2020740.
W.Zang acknowledges the support
from the Harvard-Smithsonian Center for Astrophysics through the CfA Fellowship. 
% Clement Ranc
C.R. was supported by the Research fellowship of the Alexander von Humboldt Foundation.
\end{acknowledgements}

\end{document}